\documentclass[12pt]{article}
\usepackage{amsfonts}
\usepackage{epsfig}

\newcommand{\AmS}{{\protect\the\textfont2
  A\kern-.1667em\lower.5ex\hbox{M}\kern-.125emS}}

\newcommand{\reff}[1]{(\ref{#1})}
\def\smfrac#1#2{{\textstyle\frac{#1}{#2}}}

\newcommand{\be}{\begin{equation}}
\newcommand{\ee}{\end{equation}}
\newcommand{\<}{\langle}
\renewcommand{\>}{\rangle}

\def\spose#1{\hbox to 0pt{#1\hss}}
\def\ltapprox{\mathrel{\spose{\lower 3pt\hbox{$\mathchar"218$}}
 \raise 2.0pt\hbox{$\mathchar"13C$}}}
\def\gtapprox{\mathrel{\spose{\lower 3pt\hbox{$\mathchar"218$}}
 \raise 2.0pt\hbox{$\mathchar"13E$}}}

\begin{document}
\title{Crossover scaling from classical to non-classical critical 
behaviour}
\author{
  \\
  { Sergio Caracciolo}             \\[-0.2cm]
  {\small\it Scuola Normale Superiore and INFN -- Sezione di Pisa}  \\[-0.2cm]
  {\small\it I-56100 Pisa, ITALIA}          \\[-0.2cm]
  \\[-0.1cm]  \and
  { Maria Serena Causo}             \\[-0.2cm]
  {\small\it H\"ochstleistungsrechenzentrum (HLRZ)}  \\[-0.2cm]
  {\small\it Forschungszentrum J\"ulich}  \\[-0.2cm]
  {\small\it D-52425 J\"ulich, GERMANY}          \\[-0.2cm]
  \\[-0.1cm]  \and
  { Andrea Pelissetto}    \\[-0.1cm] { Paolo Rossi}    \\[-0.1cm] { Ettore Vicari}
                          \\[-0.2cm]
  {\small\it Dipartimento di Fisica and INFN -- Sezione di Pisa}    \\[-0.2cm]
  {\small\it Universit\`a degli Studi di Pisa}        \\[-0.2cm]
  {\small\it I-56100 Pisa, ITALIA}          \\[-0.2cm]
  {\protect\makebox[5in]{\quad}}  
  \\
}
\date{}
\maketitle

\begin{abstract}
Interacting physical systems in the neighborhood of criticality 
(and massive continuum field theories) can often be characterized 
by just two physical scales: 
a (macroscopic) correlation length and a (microscopic) interaction range, 
related to the coupling and measured by the Ginzburg number $G$. 
A critical crossover limit can be defined when both scales 
become large while their ratio stays finite. 
The corresponding scaling functions are universal, 
and they are related to the standard field-theory 
renormalization-group functions. 
The critical crossover describes
the unique flow from the Gaussian to the nonclassical fixed point.

\end{abstract}

\newpage

Every physical situation of experimental relevance has at least two scales: 
one scale is intrinsic to the system, 
while the second one is related to experimental conditions.
In Statistical Mechanics (SM) the correlation length $\xi$ is 
related to experimental conditions (it depends on the temperature), 
while the interaction length (Ginzburg parameter) is intrinsic.
The opposite is true in 
Quantum Field Theory (QFT): here the correlation length (inverse mass gap) 
is intrinsic, while the interaction scale (inverse momentum) 
depends on the experiment.
Physical predictions are functions of ratios of these two scales and
describe the 
crossover from the correlation-dominated ($\xi/G$ or $p/m$ large) 
to the interaction-dominated ($\xi/G$ or $p/m$ small) regime.
In a properly defined limit they are universal and define the unique flow
between two different fixed points.

In this discussion we will consider the crossover between the Gaussian
point where mean-field predictions hold (interaction-dominated regime) 
to the standard Wilson-Fisher critical point (correlation-dominated
regime).

Massive continuum field theory is the natural setting for a 
description of this critical crossover behavior, not only in QFT, 
where only two scales characterize (super)renormalizable theories, 
but also in SM, where in principle one might expect many scales 
(lattice spacing, geometry of interactions, ...) to play a role, 
and universality may be questioned.
As we will discuss, 
critical crossover scaling exists and is universal when 
two scales become very large with respect to any other 
(microscopic) scale. Their ratio becomes the (universal) 
control parameter of the system, whose transition from $0$ 
to $\infty$ describes the critical crossover.

In recent years there has been extensive work 
\cite{Fisher-prl,A-P-K-S,%
B-B-l,B-B,B-B-M-N,A-K-S-T,B-K,L-B-B-pre,L-B-B-prl,M-B,noi-pre}
aiming at the identification of the correct, 
theoretical and experimental, definition of the critical crossover limit.
We give here a sketch of the argument for $d$-dimensional $N$-component 
vector spin models, but the notion may easily be extended to 
many other physical systems.

Let us start with the standard Landau-Ginzburg Hamiltonian on a 
$d$-dimensional lattice,
\begin{eqnarray}
&& \hskip -0.7truecm 
  {\cal H}\; =\; \sum_{i,j} \smfrac{1}{2}J(\vec{x}_i-\vec{x}_j)
\left[\phi(\vec{x}_i) - \phi(\vec{x}_j)\right]^2
\nonumber \\
&& \hskip -0.7truecm
\quad +\sum_i \left[ \smfrac{1}{2}  r\phi(\vec{x}_i)^2 +
{u\over 4!} \phi(\vec{x}_i)^4 - h \phi(\vec{x}_i)\right] ,
\label{lham}
\end{eqnarray}
where $\phi(\vec{x}_i)$ are $N$-dimensional vectors. 
We will first consider the short-range case in which 
$J(\vec{x})$ is the standard nearest-neighbour
coupling. For this model the interaction scale is controlled by the 
coupling $u$ and the relevant parameters are the (thermal) Ginzburg number
$G$ and its magnetic counterpart $G_h$ defined by:
\be
G_{\hphantom{h}} =\, u^{2/(4-d)}, \qquad
G_h =\, u^{(d+2)/[2(4-d)]}.
\ee
Under a renormalization-group (RG) transformation $G$ 
scales like the (reduced) temperature, 
while $G_h$ scales as the magnetic field. For 
$t \equiv r - r_c \ll G$ and $h\ll G_h$ one observes the standard critical 
behaviour, while in the opposite case the behaviour is classical.
The critical crossover limit corresponds to considering 
$t,h,u\to 0$ keeping $\widetilde{t} = t/G$ and $\widetilde{h} = h/G_h$ 
fixed. This limit is universal, i.e. independent of the detailed 
structure of the model: for Hamiltonians \reff{lham} 
the same behaviour is obtained as long as the interaction is 
short-ranged, i.e. for any $J(\vec{x})$ such that 
$\sum_{x}x^2\, J(\vec{x}) < + \infty$.
The crossover functions can be computed
in the standard continuum $\phi^4$ theory
\cite{B-B-l,B-B,B-B-M-N}.
A dimensional analysis shows that
(using the subtracted bare mass and removing the cutoff)
finite results can be  obtained directly in terms of the dimensionless
variable $u/t^{2-d/2} =\widetilde{t}^{d/2-2}$, and no further
limiting procedure is required. 
It is important to observe that the critical crossover functions 
are related to the standard continuum RG functions if one 
expresses them in terms of the zero-momentum 
four-point renormalized coupling $g$.
The crossover functions are well studied \cite{B-B-l,B-B,B-B-M-N}
in the fixed-dimension expansion when $d=3$.

Let us now consider the long-range case. We assume that $J(\vec{x})$
has the following form
\be
J(\vec{x})=\, \cases{ J & \qquad for $\vec{x}\in {D}$, \cr
                      0 & \qquad for $\vec{x}\not\in {D}$,
                    }
\label{defJ}
\ee
where ${D}$ is a lattice domain characterized by some scale $R$.
Explicitly we define $R$ and the corresponding domain volume
$V_R$ by
\be
V_R \, \equiv  \sum_{\vec{x}\in D} 1,
\qquad
R^2 \, \equiv {1\over 2d\,V_R} \sum_{\vec{x}\in D} x^2\; .
\label{defR}
\ee
The shape of ${D}$ is irrelevant for our purposes as 
long as $V_R\sim R^d$ for $R\to\infty$. The constant $J$ defines the 
normalization of the fields. In our discussion it is useful to 
assume a long-range normalization, i.e. $J=1/V_R$, since with this 
choice the limit $R\to\infty$ is well-defined. Notice the this
is not the normalization that is commonly used discussing
short-range models. Indeed, in the latter case, one defines
$J = R^{-2}/V_R$, so that the propagator behaves as $k^2$ for $k\to 0$.
To understand the connection between the theory with long-range
interactions and the short-range model let us perform an
RG transformation \cite{L-B-B-prl}. Define new (``blocked") coordinates 
$y_i = x_i/R$ and rescale the fields according to 
\be
\widehat{\phi}(y_i) = R^{d/2} \phi(R y_i), \qquad
\widehat{h}(y_i) = R^{d/2} h(R y_i).
\ee
The rescaled Hamiltonian becomes 
\begin{eqnarray}
&& \hskip -0.7truecm 
  \widehat{\cal H}\; =\; 
     \sum_{i,j} \smfrac{1}{2}\widehat{J}(\vec{y}_i-\vec{y}_j)
\left[\widehat{\phi}(\vec{y}_i) - \widehat{\phi}(\vec{y}_j)\right]^2
\nonumber \\
&& \hskip -0.7truecm
\quad +\sum_i \left[ \smfrac{1}{2}  r\widehat{\phi}(\vec{y}_i)^2 +
{1\over 4!} {u\over R^d}\,
    \widehat{\phi}(\vec{y}_i)^4 - 
    \widehat{h} \widehat{\phi}(\vec{y}_i)\right] ,
\label{lham1}
\end{eqnarray}
where now the coupling $\widehat{J}(\vec{x})$ is of short-range type,
i.e. independent of $R$. Being short-ranged, we can apply the 
previous arguments and define Ginzburg parameters: 
\begin{eqnarray}
&& \hskip -0.7truecm
   G_{\hphantom{h}} =  \left(u R^{-d}\right)^{2/(d-4)} = \;
    u^{2/(d-4)} R^{-2d/(4-d)}, 
\\
&& \hskip -0.7truecm
   G_h = \; R^{-d/2} \left(uR^{-d}\right)^{(d+2)/[2(d-4)]} 
\nonumber  \\ 
&& \hskip -0.7truecm \hphantom{G_h}
    = \; u^{(d+2)/[2(d-4)]}\, R^{-3d/(4-d)}.
\end{eqnarray}
Therefore, in the long-range model, the critical crossover limit can 
be defined as $R\to\infty$, $t,h\to 0$, with
$\widetilde{t}\equiv t/G$,
$\widetilde{h}\equiv t/G_h$ fixed. 
The variables that are kept fixed are the same, but a different mechanism 
is responsible for the change of the Ginzburg parameters:
in short-range models we vary $u$ keeping the range $R$ fixed and finite,
while here we keep the interaction strength $u$ fixed and vary the 
range $R$.
The important consequence of the argument presented above is that the 
critical crossover functions defined using the long-range Hamiltonian
and the previous limiting procedure agree with those computed in the
short-range model, apart from trivial rescalings.

\begin{figure}
\vspace*{-1.3cm} \hspace*{-0.5cm}
\begin{center}
\epsfxsize = 0.95\textwidth
\vspace{-2.0cm}
\leavevmode\epsffile{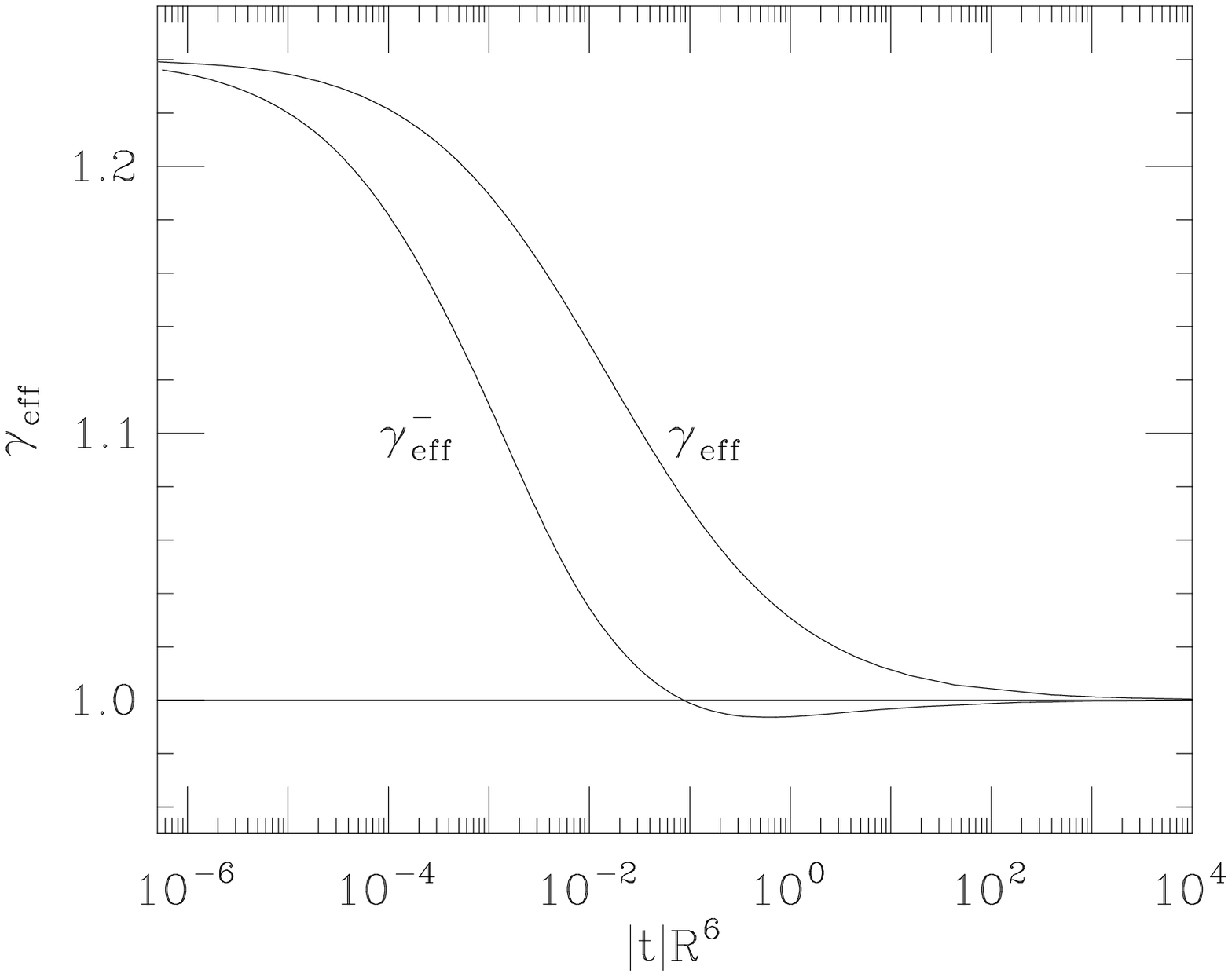}
\end{center}
\vspace*{-0.8cm}
\caption{Effective susceptibility exponent as a function of
$\widetilde{t}$ for the high- ($\gamma_{\rm eff} $) and
low- ($\gamma_{\rm eff}^- $) temperature phase of the three-dimensional
Ising model. 
}
\label{fig}
\end{figure}

Let us give a few examples. Let us introduce magnetic susceptibility,
correlation length and magnetization in the usual way:
\begin{eqnarray}
\chi &=& \! \sum_{x} \< \phi_0 \cdot \phi_x\> , \\
\xi^2 &=& \! {1\over 2d \chi} \sum_{x} x^2\, \< \phi_0 \cdot \phi_x\> , \\
M &=& \< \phi_0 \>.
\end{eqnarray}
Then one can show that in the limit $t\to 0$, $G,G_h \to 0$ with 
$\widetilde{t}$ and $\widetilde{h}$ fixed the following 
rescaled quantities have a finite limit:
\begin{eqnarray}
\widetilde{\chi} &\equiv& \chi G \to F_\chi(\widetilde{t},\widetilde{h}), 
\label{chicrossover}
\\
\widetilde{\xi}^2 &\equiv& R^{-2} \xi^2 G 
         \to F_{\xi^2}(\widetilde{t},\widetilde{h}), 
\label{xicrossover}
\\
\widetilde{M} &\equiv& M G/G_h \to F_M(\widetilde{t},\widetilde{h}),
\end{eqnarray}
where $F_\chi(\widetilde{t},\widetilde{h})$, 
      $F_{\xi^2}(\widetilde{t},\widetilde{h})$, and
      $F_M(\widetilde{t},\widetilde{h})$
are universal apart from a rescaling of $\widetilde{t}$ and 
$\widetilde{h}$ and of the functions themselves. Comparison with 
experimental data is usually performed introducing effective exponents.
For instance, for $\widetilde{h} = 0$, we can define
\begin{eqnarray}
\gamma_{\rm eff} (\widetilde{t}) &=& \! -
    \widetilde{t} {d\over d\widetilde{t}}\log F_\chi(\widetilde{t},0), 
\label{defgamma} \\
\nu_{\rm eff} (\widetilde{t}) &=& \! -
    {\widetilde{t}\over2} \;
    {d\over d\widetilde{t}}\log F_{\xi^2}(\widetilde{t},0). 
\label{defnu}
\end{eqnarray}
These functions can be related to the standard RG functions 
if one expresses them in terms of the zero-momentum four-point
coupling $g$. In the high-temperature phase one finds for instance 
(cf. Ref.~\cite{B-B})
\begin{eqnarray}
&&{\gamma_{\rm eff}(g)\over\nu_{\rm eff}(g)}=
{\gamma(g)\over\nu(g)}, \\
&&\nu(g)
  \beta(g) {d\gamma_{\rm eff}\over dg} = \gamma(g)- \gamma_{\rm eff}(g),
\end{eqnarray}
where $\gamma(g)$, $\nu(g)$, and $\beta(g)$ are the standard RG functions.
These effective exponents interpolate between the classical and the 
non-classical value. As an example, in Fig. \ref{fig}, we report the 
graph of $\gamma_{\rm eff} (\widetilde{t})$ in the high- and low-temperature
phase for the Ising model (the computation has been done using the 
results of Refs. \cite{B-B-l,B-B,B-B-M-N}). These curves are in good agreement
with numerical results for the long-range Ising model 
\cite{LB-3d}, even for very small values of $R$, i.e. for interactions
extending over a few lattice spacings.

The ideas we have presented here can be explicitly checked in the large-$N$
limit. All the crossover functions can be computed in the whole
$(t,h)$-plane for $2< d < 4$ and universality (model independence)
can be explicitly checked. For instance for the effective exponents
defined in Eqs. \reff{defgamma} and \reff{defnu} one finds in three
dimensions
\be
\gamma_{\rm eff}(\widetilde{t})\, =\, 
2\nu_{\rm eff}(\widetilde{t})\, =\; 
1 + (1 + c_\gamma \widetilde{t})^{-1/2},
\ee
where $c_\gamma$ is a non-universal constant.

One can also study the corrections to the leading universal behaviour.
If $R$ is chosen as in Eq. \reff{defR}, one verifies that the corrections to
the universal crossover functions scale as $1/R^d$ (for generic choices
of scale one would observe instead $1/R$-corrections). 

The discussion of these non-universal effects can be extended to all
values of $N$ considering a perturbative expansion around the mean-field
solution. Consider for instance the long-range Ising model
\be
{\cal H}=\, - {N\beta\over 2}
     \sum_{i,j} J(\vec{x}_i-\vec{x}_j)\,
           s(\vec{x}_i)\cdot s(\vec{x}_j),
\label{HNvector}
\ee
where $J(\vec{x})$ is defined in Eq. \reff{defJ}. For $N=1$ and for 
a particular choice of $D$ this is the model that has been studied 
in Refs. \cite{L-B-B-pre,L-B-B-prl,LB-3d}. Computing the corrections to
mean field allows us to determine the corrections to $\beta_c(R)$ for 
$R\to\infty$ for the Hamiltonian \reff{HNvector}. One finds for $d>2$
\be
\beta_c(R) =\; \int {d^dk\over (2\pi)^d} {1\over \Pi(k)}
\ee
with corrections of order $R^{-2d}$ (
multiplicative logarithms appear
for some special values of $d$, for instance for $d=3$). Here
\be
\Pi(k) =\, {1\over V_R}\, \sum_{x\in D}\left(1 - e^{ikx}\right),
\ee
and the integral is extended over the first Brillouin zone.
Expanding the integral in powers of $R$ one finds
\be
\beta_c(R) =\; 1 + {\alpha\over R^d} + \ldots 
\ee
where $\alpha$ depends on the precise definition of the domain $D$. In two
dimensions there are logarithmic corrections and one finds
\be
\beta_c(R) =\; 1 +\, 
   {1\over 4\pi R^2} \log R^2 +\; O(R^{-2}).
\ee
By considering the mean-field limit one can also relate the non-universal
constants that appear in the definition of the crossover scaling functions.
To give an example, consider $\widetilde{\chi}$ for $\widetilde{h}=0$.
The function $F_\chi(\widetilde{t},0)$ can be computed in perturbation
theory, cf. Ref. \cite{B-B}, obtaining a function 
$F_\chi^{\rm sr}(\widetilde{t})$. On the other hand, if one considers
the model \reff{HNvector} one obtains a different 
$F_\chi^{\rm lr}(\widetilde{t})$. As we explained above we should have
\be
F_\chi^{\rm lr}(\widetilde{t}) = \; a_\chi\,
         F_\chi^{\rm sr}(b \widetilde{t})\; .
\ee
The analysis of the mean-field limit provides exact expressions
for $a_\chi$, which depends on the observable, and $b$, which 
depends only on the model. 

Finally we want to discuss the crossover scaling limit for models
that have $\beta_c = +\infty$. This is the case of the 
two-dimensional $N$-vector model with $N\ge 3$. For these
theories define
\be
\widehat{t}\equiv 
  R^2\left(1 + {1\over 4\pi R^2} \log R^2 -\, \beta\right)
\ee
and consider the limit $R\to\infty$ with $\widehat{t}$ fixed.
One finds that the limits defined in 
Eqs. \reff{chicrossover} and \reff{xicrossover} still exist 
and define crossover functions of $\widehat{t}$. 
For $\widehat{t}\to+\infty$ these functions show
mean-field behaviour, while standard asymptotic scaling 
is observed for $\widehat{t}\to-\infty$.
Notice that one can use $\widehat{t}$ as a variable instead of 
$\widetilde{t}$ also when $\beta_c$ is finite. In this case,
however, nothing new is obtained, since 
$\widehat{t} - \widetilde{t}$ is simply a constant for $R\to \infty$.

The model with Hamiltonian \reff{HNvector} can be studied in the limit 
$N\to 0$. In this case it can be rewritten in terms of 
self-avoiding walks (SAWs) \cite{Daoud-etal_75} with long-range jumps.
To be explicit, we define an $n$-step SAW with range $R$ as a 
sequence of lattice
points $\{\omega_0,\cdots,\omega_n\}$ with $\omega_0 = (0,0,0)$ and
$\omega_{j+1} \in D_R(\omega_j)$, such that
$\omega_i\not=\omega_j$ for all $i\not= j$. 
Then, if $c_{n,R}(x)$ is
the number of $n$-step SAWs with range $R$ going from $0$ to $x$,
we indicate with $c_{n,R}$
the total number of $n$-step walks and with $E^2_{n,R}$
the mean square end-to-end distance. They are defined as:
\begin{eqnarray}
c_{n,R} &=& \sum_x c_{n,R}(x), \\
E^2_{n,R} &=& {1\over c_{n,R}} \sum_x x^2 c_{n,R}(x).
\end{eqnarray}
One can then prove that
\begin{eqnarray}
\lim_{N\to 0} \chi(\beta) &= &
     \sum_{n=0}^\infty \widehat{\beta}^n c_{n,R} \; ,
\label{chiRNeq0} \\
\lim_{N\to 0}  \xi^2(\beta) \chi(\beta) &= & {1\over 2d}
    \sum_{n=0}^\infty \widehat{\beta}^n c_{n,R} E^2_{n,R}.
\label{xiRNeq0}
\end{eqnarray}
where $\widehat{\beta} = \beta/V_R$ and $\chi$ and $\xi^2$ are defined 
in the model \reff{HNvector}.

The crossover limit is trivially defined remembering that $n$
is the dual variable (in the sense of Laplace transforms) of
$t$. Therefore we should study the limit
$n\to\infty$, $R\to\infty$ with $\widetilde{n}\equiv n R^{-2d/(4-d)}$ fixed.
From Eqs. \reff{chicrossover} and 
\reff{xicrossover} we obtain that the following limits
exist:
\begin{eqnarray}
\widetilde{c}_{n,R} \equiv \;
c_{n,R} \beta_{c}(R)^n &\to& g_c(\widetilde{n}), \\
\widetilde{E}^2_{n,R} \equiv \;
E^2_{n,R} R^{-8/(4-d)} &\to &g_E(\widetilde{n}),
\end{eqnarray}
where the functions $g_c(\widetilde{n})$ and $g_E(\widetilde{n})$ are related
by a Laplace transform to $F_\chi(\widetilde{t},0)$ and
$F_{\xi^2}(\widetilde{t},0)$. Explicitly
\begin{eqnarray}
F_\chi(t,0) &=& \int^\infty_0 du\,g_c(u) e^{-ut},  \\
F_{\xi^2}(t,0) F_\chi(t,0) &=&  
{1\over 2d} \int^\infty_0 du\, g_c(u) g_E(u) e^{-ut}. 
\nonumber \\ [-2mm]
{}
\end{eqnarray}
Using perturbation theory it is possible to derive predictions for $E^2_n$
and $c_n$. For $E^2_n$ we can write
\be
g_{E,PT}(\widetilde{n}) =\; a_E\, \widetilde{n}\, h_E(z),
\label{gEPT}
\ee
where $z = (\widetilde{n}/l)^{1/2}$. The function 
$h_E(z)$ has been computed in perturbation theory to six-loop order
\cite{Muthukumar-Nickel}. Resumming the series with a Borel-Leroy
transform one finds that a very good approximation is provided
by \cite{Belohorec-Nickel_97}
\be
h_E(z) = (1 + 7.6118 z + 12.05135 z^2)^{0.175166}\; .
\label{hEresum}
\ee
Comparison with a detailed Monte Carlo simulation for the
Domb-Joyce model indicates \cite{Belohorec-Nickel_97}
that this simple expression
differs from the exact result by less than 0.02\% for $z < 2$.

The constants $a_E$ and $l$ appearing in Eq. \reff{gEPT}
are non universal. For our specific model they are given by
\be
a_E = 6,\qquad l =\, (4\pi)^3.
\ee
We have performed \cite{crossoverSAW} 
an extensive simulation of this model of long-range SAWs 
generating walks 
of length up to $N\approx 7 \cdot 10^4$. The domain $D$ was chosen
as follows:
\be
D = \; \left\{x:\, \sum_i |x_i|\le \rho\right\}\; .
\ee
In the simulation we varied $\rho$ between 2 and 12.
Let us describe the results for the end-to-end distance. 
Analogous results can be obtained for $c_{n,R}$.

\begin{figure}
\vspace*{-0.0cm} \hspace*{0cm}
\begin{center}
\epsfxsize = 0.95\textwidth
\vspace{-2.0cm}
\leavevmode\epsffile{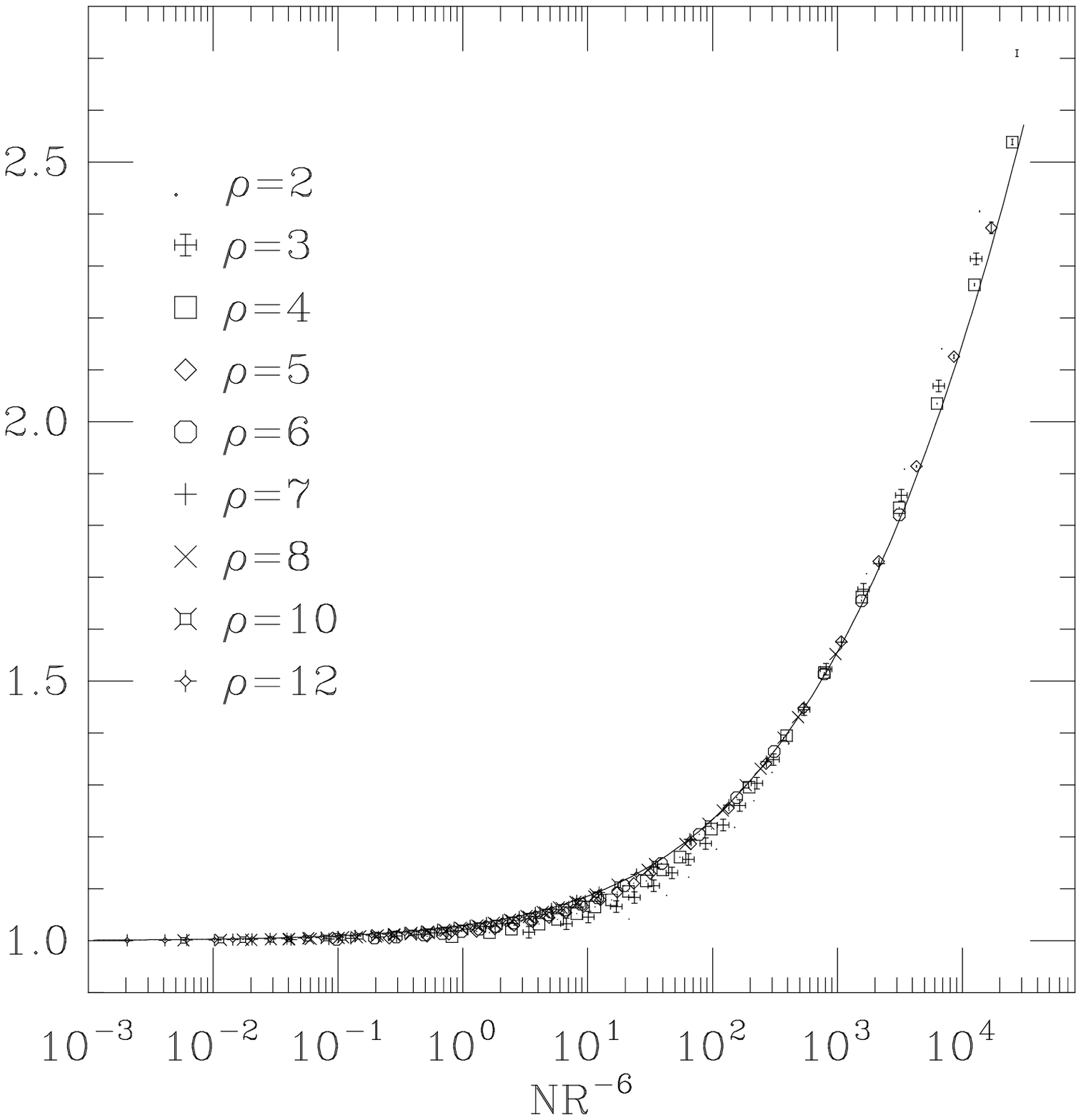}
\end{center}
\vspace*{-0.8cm}
\caption{Results for $\widetilde{E}^2_R/(6\widetilde{n})$. 
The solid line is the 
theoretical prediction (\protect\ref{gEPT}),
(\protect\ref{hEresum}).
}
\label{figR2}
\end{figure}

\begin{figure}
\vspace*{-0.0cm} \hspace*{0cm}
\begin{center}
\epsfxsize = 0.95\textwidth
\vspace{-2.0cm}
\leavevmode\epsffile{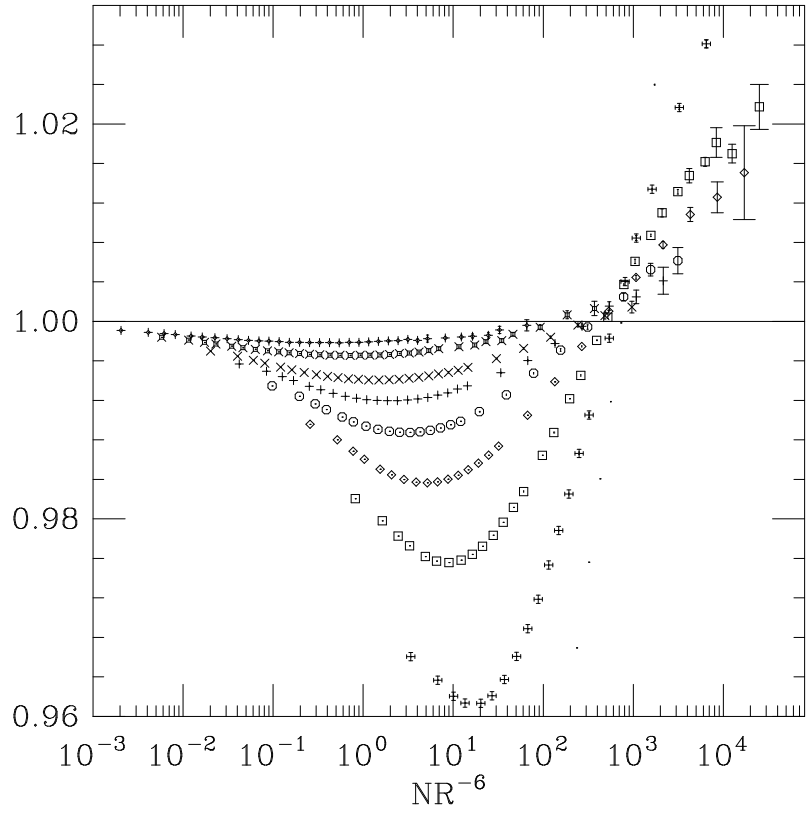}
\end{center}
\vspace*{-0.8cm}
\caption{Results for $\widetilde{E}^2_R/g_{E,PT}(\widetilde{n})$. 
We use the same symbols as in Fig. \protect\ref{figR2}.
}
\label{figratio}
\end{figure}

In Fig. \ref{figR2} we report our results for $\widetilde{E}^2_{n,R}$ together
with the perturbative prediction $g_{E,PT}(\widetilde{n})$
defined in Eqs. (\ref{gEPT},\ref{hEresum}).
The agreement is very good although one can see clearly the presence
of corrections to scaling. In order to see better the discrepancies between
the numerical data and the theoretical prediction we report in Fig. 
\ref{figratio}
the ratio $\widetilde{E}^2_{n,R}/g_{E,PT}(\widetilde{n})$. In this plot the
corrections to scaling are clearly visible: points with different $R$ fall on
different curves that converge to 1 as expected. For $\rho = 12$, the
deviations are less than 0.2\%. It is interesting to observe that
the corrections change sign with $\widetilde{n}$. For
$\widetilde{n} \ltapprox 2 \cdot 10^2$, the corrections are negative, while
in the opposite case they are positive.

\begin{figure}
\vspace*{-0.0cm} \hspace*{0cm}
\begin{center}
\epsfxsize = 0.95\textwidth
\vspace{-2.0cm}
\leavevmode\epsffile{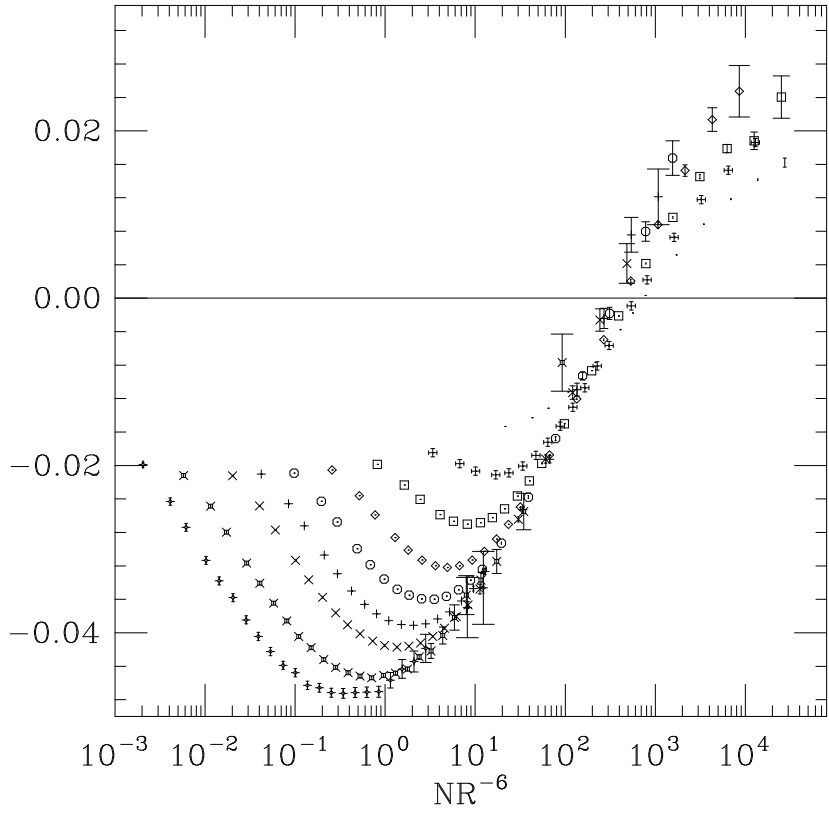}
\end{center}
\vspace*{-0.8cm}
\caption{Results for $\Delta_{E;n,R}$.
We use the same symbols as in Fig. \protect\ref{figR2}.
}
\label{figdelta}
\end{figure}

The corrections to scaling are expected\footnote{This behaviour
should be observed only for quantities that are defined
using $R$ as scale. If we were considering for instance
$E^2_{n,R} \rho^{-8/(4-d)}$ we would of course obtain the same universal
limiting curve, but now with corrections of order $1/\rho$.}
to scale as $R^{-d}$. To check this behaviour let us consider
\be
\Delta_{E;n,R}\equiv
 \left[\widetilde{E}^2_{n,R}/g_{E,PT}(\widetilde{n}) - 1\right]\, R^3\;.
\ee
The plot of $\Delta_{E;n,R}$ is reported in Fig. \ref{figdelta}. A good scaling
behaviour is observed confirming the theoretical prediction for the
corrections. This nice scaling indicates also
that the approximation \reff{hEresum} can be considered
practically exact at our level of precision.

\begin{figure}
\vspace*{-0.0cm} \hspace*{0cm}
\begin{center}
\epsfxsize = 0.95\textwidth
\vspace{-2.0cm}
\leavevmode\epsffile{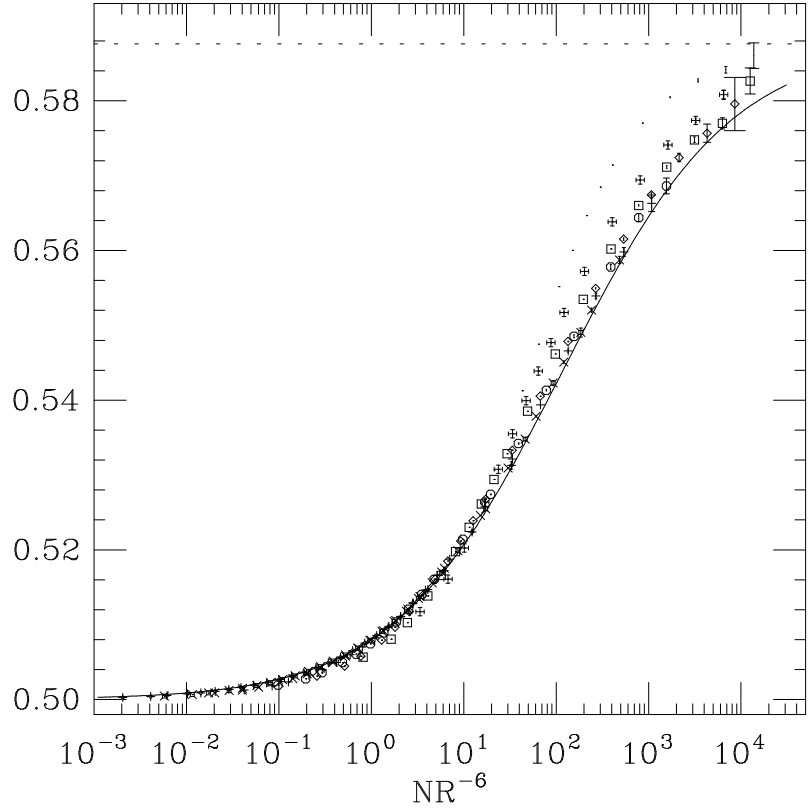}
\end{center}
\vspace*{-0.8cm}
\caption{Results for $\nu_{\rm eff}$. 
We use the same symbols as in Fig. \protect\ref{figR2}.
The dashed line is the self-avoiding walk value $\nu=0.58758(7)$.
The solid line is the theoretical prediction.
}
\label{fignu}
\end{figure}

We have also defined an effective exponent $\nu_{\rm eff}$
\be
\nu_{\rm eff} =\, {1\over 2\log 2}\; 
\log\left( {E^2_{2n,R}\over E^2_{n,R}} \right)\; .
\ee
It is reported in Fig. \ref{fignu}. It shows the expected
crossover behaviour between the mean-field value $\nu = 1/2$
and the self-avoiding walk value $\nu = 0.58758(7)$ 
\cite{Belohorec-Nickel_97}.

\newpage

\end{document}